\documentclass[twocolumn,aps,prl,amssymb,]{revtex4}

\usepackage[dvips]{graphicx}
\usepackage{amsmath}
\usepackage{SIunits}
\usepackage{color}

\begin{document}

\title{Three-Photon Correlations in a Strongly Driven Atom-Cavity System}
\author{Markus Koch}
\email{markus.koch@mpq.mpg.de}
\author{Christian Sames}
\author{Maximilian Balbach}
\author{Haytham Chibani}
\author{Alexander Kubanek}
\altaffiliation[Present address: ]{Department of Physics, Harvard University, Cambridge, Massachusetts, 02138, USA}
\author{Karim Murr}
\author{Tatjana Wilk}
\author{Gerhard Rempe}
\affiliation{Max-Planck-Institut f\"ur Quantenoptik, Hans-Kopfermann-Str. 1,
D-85748 Garching, Germany}

\date{\today}

\begin{abstract}
The quantum dynamics of a strongly driven, strongly coupled single-atom-cavity system is studied by evaluating time-dependent second- and third-order correlations of the emitted photons. The coherent energy exchange, first, between the atom and the cavity mode, and second, between the atom-cavity system and the driving laser, is observed. Three-photon detections show an asymmetry in time, a consequence of the breakdown of detailed balance. The results are in good agreement with theory and are a first step towards the control of a quantum trajectory at larger driving strength.
\end{abstract}


\maketitle
Open quantum systems far from thermal equilibrium hold great promise for the investigation of fundamental physics and the implementation of practical devices~\cite{Mabuchi02,Kimble08}. The versatility of such systems comes from two features: the coherent evolution induced by the driving and the dissipation enabling a transfer of information to an observer. These two characteristics affect each other, and the deterministic evolution is interrupted by unpredictable quantum jumps~\cite{Carmichael93,Molmer93}. Monitoring such a quantum trajectory is a challenge, in particular when many quantum states must be discriminated from each other. A model system in this context is provided by optical cavity quantum electrodynamics (QED) in the regime of strong light-matter coupling, where atomic and photonic observables have been tracked in real time~\cite{Hood00,Pinkse00,Foster00,Khudaverdyan09} and controlled by means of feedback~\cite{Smith02,Kubanek09,Koch10}. However, these experiments were performed at low excitation. Stronger driving and, hence, faster probing would allow one to track the system more closely and explore high-intensity effects like the coherent coupling of the system with the drive laser or the dynamical polarization of the dressed states~\cite{Alsing91,Armen09}. Moreover, higher excited states containing several photons should be discernible by characteristic patterns of multiple-photon emissions~\cite{Kubanek08}, which can be asymmetric in time due to the predicted breakdown of detailed balance~\cite{Denisov02}. In this Letter we explore such patterns for a strongly driven atom-cavity system, when the excitation rate exceeds the dissipative rates.

\begin{figure}
\includegraphics[width=8.6cm]{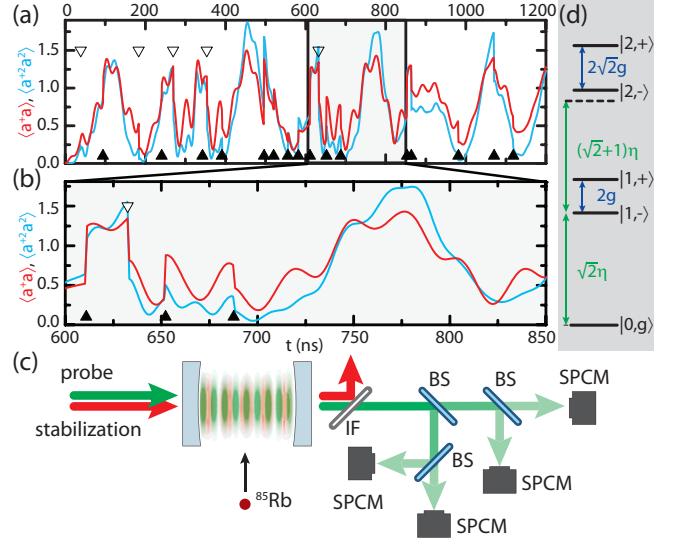}
\caption{\label{Figure1} (Color online) (a) A calculated quantum trajectory of $\langle a^{\dagger}a\rangle$ (red) and $\langle a^{\dagger2}a^{2}\rangle$ (blue). (b) A zoom reveals coherent oscillations at different frequencies accompanied by irreversible quantum jumps due to spontaneous emission ($\nabla$) and cavity decay ($\blacktriangle$). (c) Experimental setup: Single $^{85}$Rb atoms pass through the cavity mode. Probe light and stabilization light are separated using an interference filter (IF). The probe light is split by beam-splitters (BS) into four equal parts each directed onto a single-photon counting module (SPCM). (d) Energy levels of the strongly coupled atom-cavity system with the dressed eigenstates $|n,\pm\rangle=(|n,g\rangle\pm|n-1,e\rangle)/\sqrt{2}$ and the coupling (green arrows) induced by the probe laser.}
\end{figure}

We consider a system where the atom-cavity coupling strength $g_0$ exceeds the atomic polarization decay rate $\gamma$ and the cavity field decay rate $\kappa$. The internal dynamics, described by the Jaynes-Cummings Hamiltonian, is complemented by driving with a probe laser of strength $\eta$~\footnote{Driving Hamiltonian: $H_{d}=\hbar\eta(a+a^\dagger)$ with photonic annihilation and creation operators $a$ and $a^\dagger$, respectively.} and dissipation due to spontaneous emission and cavity decay. By monitoring the photon stream from the cavity, one can evaluate different observables such as the average photon number, $\langle a^{\dagger} a\rangle$, or the average number of photon pairs, $\langle a^{\dagger2} a^{2}\rangle$. The first is interesting, e.g., in the context of normal-mode spectroscopy~\cite{Boca04,Maunz05}, while insight into quantum effects can be obtained by regarding photon pairs~\cite{Schuster08,Kubanek08}. Due to the interplay of the different dynamical processes, both observables are expected to undergo complex dynamics as illustrated by the calculated quantum trajectory depicted in Fig.~\ref{Figure1}\,(a) and (b), cf. also~\cite{Tian92}.
It displays coherent oscillations at different frequencies interrupted by sudden quantum jumps due to spontaneous emission ($\nabla$) and cavity decay ($\blacktriangle$). In the following, the time evolution of $\langle a^{\dagger} a\rangle$ conditioned upon such a quantum jump is studied experimentally by evaluating the time-dependent second-order intensity correlation function. In addition, we introduce measurements of the third-order intensity correlation function as a new tool to study the time-dependence of $\langle a^{\dagger2} a^{2}\rangle$ and to investigate the dynamics of the system conditioned upon two simultaneous or successive detection events. While second-order correlations are naturally symmetric in time, third-order correlations enable us to address the simple, but experimentally unexplored, question whether the emitted photon stream is symmetric in time.

The experimental setup is depicted in Fig.~\ref{Figure1}\,(c) and has been described before~\cite{Koch10}. A Fabry-Perot resonator with a finesse of 195000 and a length of 260\,$\mu$m, yielding $\kappa/2\pi$\,=\,1.5\,MHz, is tuned to the cycling transition from $F$\,=\,3, $m_{F}$\,=\,3 to $F$\,=\,4, $m_{F}$\,=\,4 of the D$_2$-line of $^{85}$Rb with $\gamma/2\pi$\,=\,3\,MHz. The atom-cavity coupling to the TEM$_{00}$ mode with maximal coupling strength $g_{0}/2\pi$\,=\,16\,MHz puts the experiment well into the strong coupling regime. A laser at 785\,nm, detuned by four free-spectral ranges, is used to stabilize the cavity length such that the bare atom is on resonance with the cavity mode. The coupled system is driven by a circularly polarized probe laser at 780\,nm. In order to excite higher-order dressed states, cf. Fig.~\ref{Figure1}\,(d), and to increase the signal, we use relatively strong probe powers. Since trapping atoms is rather difficult in this parameter regime, we perform the measurements with atoms passing through the mode, launched via an atomic fountain from underneath the cavity. The transit time of an individual atom is about 20\,$\mu$s. The attractive potential induced by the stabilization light at 785\,nm (AC-Stark shift 5\,MHz) guides the passing atoms towards regions of strong coupling. With a detuning of the probe laser with respect to the cavity of $\Delta_c/2\pi$\,$=$\,-12\,MHz, near-resonant with the lower-frequency normal mode, cf. Fig.~\ref{Figure1}\,(d), a transient atom causes an increase of the probe light transmission. We evaluate the recorded photon clicks only in those intervals where the transmission is increased by at least a factor of 1.6 compared to the empty cavity value. In each launch about 25 atoms cause such an increase. In the presence of an atom, the probability of having a second atom in the cavity is less than 3\,$\%$.

First, we consider the time evolution of the average photon number shown as a red line in Fig.~\ref{Figure1}\,(a) and (b). Two characteristic frequencies are visible, a slow oscillation with a period around 150~ns and a fast oscillation with a period of about 30~ns. To observe these in the experiment, we evaluate the second-order correlation function, $g^{(2)}(\tau)=\langle a^\dagger a^\dagger(\tau)a(\tau)a\rangle/\langle a^\dagger a\rangle^{2}$, of the transmitted probe light measuring the conditional evolution of the average photon number after the detection of a photon.

\begin{figure}
\includegraphics[width=8.6cm]{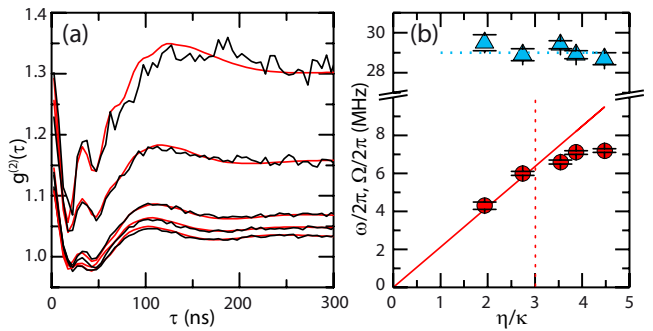}
\caption{\label{powerdep} (Color online) (a) $g^{(2)}(\tau)$ for different driving strengths (from top to bottom $\eta=1.9\kappa,~2.7\kappa,~3.5\kappa,~3.9\kappa,~4.5\kappa$) in the experiment (black) and theory (red). (b) The two oscillation frequencies $\omega$ ($\triangle$) and $\Omega$ ($\circ$) as a function of the driving strength. $\omega$ is constant with an average $\bar{\omega}/2\pi$\,=\,29\,MHz (dashed blue line) close to the maximum vacuum Rabi frequency of $2g_{0}/2\pi$\,=\,32\,MHz. $\Omega$ increases with $\eta$. Up to a driving strength of 3$\kappa$ (dashed red line) it can be described by a two-level model with $\Omega$\,=\,$\sqrt{2}\eta$ (solid red line).}
\end{figure}

Experimental results are plotted in Fig.~\ref{powerdep}\,(a) for different driving strengths (black). Deviations of the asymptotic values from 1 result from the variation of the transmitted intensity during the passage of an atom through the mode and are well understood~\cite{Muenstermann99,Alton2011}. Also shown are calculations (red), for which details can be found in the supplementary information. Since the atomic transit happens on a much longer timescale than the internal dynamics, we can account for it in the theory by averaging the correlation function over a proper atomic position distribution, which is the same for all calculations in this work. After an additional vertical scaling of the theoretical curves by up to 10~$\%$ to match the asymptotic values of the experiment, we find good agreement between theory and experiment.

We identify two different oscillation frequencies in the correlation functions. For a quantitative analysis, we fit an exponentially damped, oscillating function~\footnote{$f(\tau)$\,=\,$e^{-\tau/T} (A_{\omega} \cos(\omega \tau) - A_{\Omega} \cos(\Omega\tau-\phi_{\Omega})) + f_{0}$.}. The obtained frequencies are plotted in Fig.~\ref{powerdep}\,(b) as a function of the driving strength $\eta$. The faster oscillation frequency $\omega$ ($\blacktriangle$) reflects the coherent exchange of energy between the cavity mode and the atom, i.e. the vacuum Rabi oscillations~\cite{Rempe91,Bochmann08}. We find an almost constant frequency with an average of $\bar{\omega}/2\pi$\,=\,29\,MHz (dashed blue line). Due to the atomic motion this is slightly smaller than the maximum expected value of $2g_{0}/2\pi=32$~MHz.

The strong driving gives rise to another coherent process, namely the exchange of energy between the atom-cavity system and the drive laser. This results in another characteristic oscillation frequency $\Omega$ ($\bullet$) which depends on the driving strength. As these oscillations are the dynamic manifestation of the supersplitting of the vacuum Rabi resonance~\cite{Bishop09}, we will refer to them as super Rabi oscillations. Due to the anharmonic energy-level structure of the system, cf. Fig. \ref{Figure1} (d), it behaves at low excitation like a driven two level system \cite{Tian92}. Neglecting the atom-cavity detuning due to the AC-Stark shift of the stabilization laser and the small detuning from the normal mode, we expect a Rabi frequency of $\Omega$\,=\,$\sqrt{2}\eta$, which is plotted as a solid red line in Fig.~\ref{powerdep}\,(b). Deviations from the two-level approximation occur when the transition to the second dressed state becomes important. This is expected for a driving strength exceeding 3$\kappa$, marked as a vertical dashed line. The reduction of the oscillation frequency compared to the two-level approximation at higher powers is in agreement with previous studies of a driven anharmonic oscillator~\cite{Claudon08}.

Next, we consider the dynamics of another observable, namely the probability for the emission of a photon pair, $\langle a^{\dagger2}a^{2}\rangle$, blue line in Fig.~\ref{Figure1}\,(a) and (b). This is motivated by the fact that the dynamics of the average photon number, $\langle a^{\dagger}a\rangle$, seems to be dominated by the coherent internal dynamics (vacuum Rabi oscillations) and driving (super Rabi oscillations) of the first-order dressed states only. The quantum Rabi oscillation~\cite{Brune96} at a frequency of $2\sqrt{2}g$ of the second-order dressed states are not visible. As these states emit photons in pairs~\cite{Kubanek08} the probability to detect a photon pair is more sensitive to the occupation of the second-order dressed states than the average photon number. This is supported by the corresponding quantum trajectory. Here, the expectation value $\langle a^{\dagger2}a^{2}\rangle$ also undergoes super Rabi oscillations at the same frequency as $\langle a^{\dagger}a\rangle$ but with a higher visibility. However, the fast oscillations, nicely visible in Fig.~\ref{Figure1}\,(b), clearly deviate in frequency and visibility from the vacuum Rabi oscillations that appear for $\langle a^{\dagger}a\rangle$.

To confirm this behavior experimentally, we evaluate the probability to detect a photon pair at a time $\tau$ after a single photon has been observed, corresponding to the third-order correlation function $g^{(3)}(\tau,0)=\langle a^\dagger a^\dagger(\tau)^{2}a(\tau)^{2}a\rangle/\langle a^\dagger a\rangle^{3}$~\cite{Glauber63}. For $\tau$\,$>$\,0, it measures the time dependence of $\langle a^{\dagger2}a^{2}\rangle$ conditioned upon the detection of a single photon. For $\tau$\,$<$\,0, it measures the time dependence of $\langle a^{\dagger}a\rangle$ conditioned upon the detection of a photon pair. A similar time dependence as for the second-order correlation function is expected in the latter case.

\begin{figure}
\includegraphics[width=8.6cm]{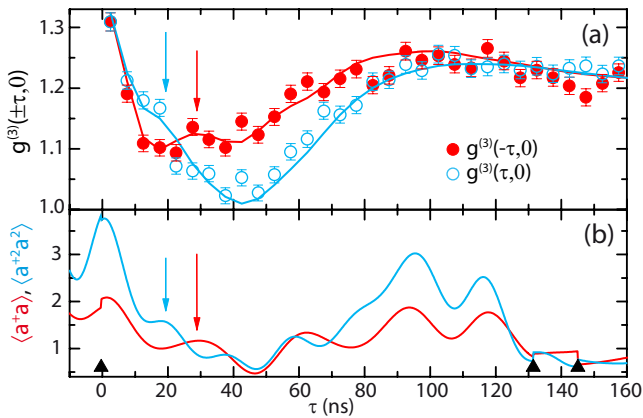}
\caption{\label{g3_zoom} (Color online) (a) $g^{(3)}(\pm\tau,0)$ ($\eta$\,=4.5\,$\kappa$). The vacuum Rabi oscillations at a frequency of $2g_{0}$ are visible (red arrow) as a weak modulation for negative times (red, $\bullet$) whereas the quantum Rabi oscillations at a frequency of $2\sqrt{2} g_{0}$ are visible (blue arrow) as a weak modulation for positive times (blue, $\circ$). (b) A sample quantum trajectory shows the same characteristic frequencies for the observables $\langle a^{\dagger}a\rangle$ (red) and $\langle a^{\dagger2}a^{2}\rangle$ (blue).}
\end{figure}

Experimental and theoretical results are plotted in Fig.~\ref{g3_zoom}\,(a) showing good agreement. It is instructive to compare these correlation functions directly to a sample quantum trajectory shown in Fig.~\ref{g3_zoom}\,(b) where $\tau$\,=\,$0$ is defined by the detection of a photon. The correlation function behaves differently for positive and negative times. On very short timescales, a weak modulation at different frequencies is clearly visible. For $\tau$\,$<$\,0, a local maximum at about 30~ns (red arrow) appears resulting from vacuum Rabi oscillations between the normal modes as observed in $g^{(2)}(\tau)$. The position of the peak matches with the oscillation period of the average photon number in the quantum trajectory which is also marked by a red arrow. For $\tau$\,$>$\,0, we find a \textit{shoulder} at about 20~ns (blue arrow) which was consistently reproduced in other measurements at different detunings and driving strengths. It is in good agreement with the oscillation frequency of $\langle a^{\dagger2}a^{2}\rangle$ in the quantum trajectory. This faster oscillation frequency is a consequence of the quantum Rabi oscillations of the second-order dressed states with an expected period of about $2\pi/2\sqrt{2}g_{0}$\,=\,22\,ns~\cite{Brune96,Hofheinz08}.

For longer times, the onset of the super Rabi oscillations is also visible in $g^{(3)}(\tau,0)$. While its frequency seems to be similar for positive and negative times, the amplitude is more pronounced in the first case. Both observations are in agreement with our previous statement, based on the sample quantum trajectory, that $\langle a^{\dagger2}a^{2}\rangle$ and $\langle a^{\dagger}a\rangle$ oscillate slowly at a similar frequency but with different visibility.

\begin{figure}
\includegraphics[width=8.6cm]{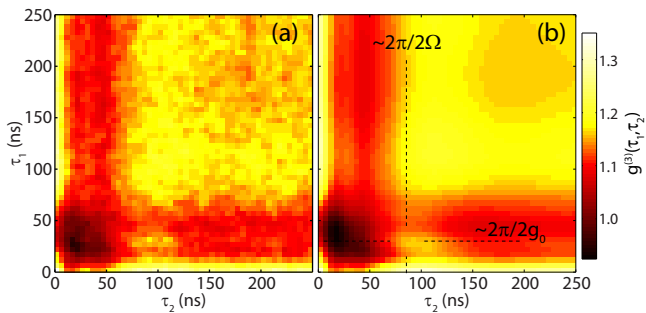}
\caption{\label{g32D} (Color online) $g^{(3)}(\tau_{1},\tau_{2})$ in the experiment (a) and theory (b) ($\eta$\,=3.9\,$\kappa$). To reduce noise, it was convoluted with a two-dimensional Gaussian filter with a width (FWHM) of 7\,ns. An intuitive explanation for the pronounced, time-asymmetric peak marked by the dashed lines is given in the text.}
\end{figure}

Finally, we investigate how patterns of two successive photon detections determine the future trajectory of the system. As an example, consider the quantum trajectory shown in Fig.~\ref{Figure1}\,(b), and here the two photons which are emitted at about 650~ns and 680~ns. We ask the question whether and how the subsequent time evolution of the photon number depends on the separation between such two photon detections. To give an answer, we evaluate the general third-order correlation function $g^{(3)}(\tau_1,\tau_2)=\langle a^\dagger a^\dagger(\tau_1) a^\dagger(\tau_1+\tau_2)a(\tau_1+\tau_2)a(\tau_1)a\rangle/\langle a^\dagger a\rangle^{3}$. It is proportional to the probability to detect three photons, with time separations $\tau_{1}$ between first and second, and $\tau_{2}$ between second and third photon. An example ($\eta$\,=\,3.9\,$\kappa$) is shown in Fig.~\ref{g32D} for experiment (a) and theory (b). The special case $g^{(3)}(\tau,0)$ discussed previously can be found on the vertical (for $\tau$$>$$0$) and horizontal (for $\tau$$<$$0$) axis of this figure. Our question can now be answered by comparing different horizontal cuts through the figure. Each cut measures the time evolution of the average photon number after the detection of a photon pair separated by $\tau_{1}$. We see immediately that there is a strong dependence on $\tau_{1}$, i.e. on the previous measurement record, if this separation is shorter than the coherence time of the system.

The most interesting feature is the occurrence of a peak marked by two dashed lines appearing at $\tau_{1}$\,$\approx$\,$2\pi/2g_{0}$ and $\tau_{2}$\,$\approx$\,$2\pi/2\Omega$, i.e. when the separation between the first two clicks is one period of a vacuum Rabi oscillation whereas the separation between the second and the third click is half a period of the super Rabi oscillations. These timescales suggest that we observe an interplay between the vacuum Rabi oscillation and the super Rabi oscillation. We give an intuitive explanation in terms of the quantum measurement process: The detection of two photons separated by $\tau_{1}$\,$\approx$\,$2\pi/2g_{0}$ is a signature of the normal modes which oscillate at this frequency. Between the two detection events the state of the system therefore had a large contribution from the first-order dressed states. The detection of the second photon projects these states onto the ground state. Subsequently, the super Rabi oscillations cause a peak of the excitation after half a super Rabi oscillation period, i.e. $\tau_{2}$\,$\approx$\,$2\pi/2\Omega$. The reverse process at $\tau_{1}$\,$\approx$\,$2\pi/2\Omega$ and $\tau_{2}$\,$\approx$\,$2\pi/2g_{0}$ is not particularly enhanced: the super Rabi frequency is not characteristic to any particular set of dressed states. Therefore, the detection of two photons separated by half a period of the super Rabi frequency does not change the subsequent time evolution much compared to the detection of just an individual photon. This explains the missing of a similar peak at $\tau_{1}$\,$\approx$\,$2\pi/2\Omega$ and $\tau_{2}$\,$\approx$\,$2\pi/2g_{0}$.

Both discussed effects, the different dynamics of the first- and second-order dressed states as well as the interplay between vacuum and super Rabi oscillations, give rise to pronounced time-asymmetries, i.e. $g^{(3)}(\tau_{1},\tau_{2})\neq g^{(3)}(\tau_{2},\tau_{1})$, demonstrating that the transmitted photon stream is asymmetric in time. This effect cannot be observed for the individual parts of the system, neither the empty cavity nor a two-level system in free space. It confirms previous evidence for an asymmetry in studies of intensity-field correlation functions~\cite{Foster00}. The occurrence of such time asymmetric fluctuations in the output fields can be considered as a direct evidence for the breakdown of detailed balance in a driven system far away from thermal equilibrium~\cite{Denisov02}.

In conclusion, using second- and third-order intensity correlations, we were able to probe the dynamics of the normal modes and the second rung of dressed states. All relevant dynamical processes - dissipation, driving and internal dynamics - have been observed. The next step will be to use the information gained by two detection events to control the system by means of feedback and stabilize its state against fluctuations. Moreover, higher-order correlation functions enable a full characterization of the system photon statistics and can therefore be used to demonstrate the non-classical nature of the higher-order dressed states~\cite{Hong10}.

We thank H.\,J.~Carmichael for helpful discussions and S.~D\"urr for comments on the manuscript. Financial support from the DFG (Research Unit 635), the EU (IST project AQUTE) and the Bavarian PhD program of excellence (QCCC) is gratefully acknowledged.

Note added: During the preparation of the manuscript we became aware of an experiment reporting on 
super Rabi oscillations in a circuit QED experiment~\cite{Lang11}.

\end{document}


\section{Master Equation}
In the rotating wave approximation, the driven Jaynes-Cummings Hamiltonian reads
\[
H=\hbar\Delta_{a}\sigma_{+}\sigma+\hbar\Delta_{c}a^{\dagger}a+\hbar g(a^{\dagger}\sigma_{-}+a\sigma_{+})+\hbar\eta(a+a^{\dagger})
\]
where $\Delta_{a}=\omega_a-\omega_l$ and $\Delta_{c}=\omega_c-\omega_l$ are the atom-laser and cavity-laser detunings, respectively, $a$ ($a^\dagger$) and $\sigma_-$ ($\sigma_+$) are the photonic and atomic annihilation (creation) operators, $g$ is the coupling strength and $\eta$ is the driving strength. In the presence of dissipation caused by spontaneous emission and cavity decay, the time evolution of the system is governed by a Lindblad master equation~\cite{Carmichael93}:
\[
\dot{\rho}=\mathcal{L}\rho =\frac{1}{i\hbar}[H,\rho]+\kappa(2a\rho a^{\dagger}-a^{\dagger}a\rho-\rho a^{\dagger}a)+\gamma(2\sigma_{-}\rho \sigma_{+}-\sigma_{+}\sigma_{-}\rho-\rho \sigma_{+}\sigma_{-})
\]
where $\gamma$ is the atomic polarization decay rate and $\kappa$ is the cavity field decay rate. The steady-state density matrix, describing the time-averaged state of the system, is defined by
\[
\mathcal{L}\rho_{ss}=0.
\]
After truncating the Hilbert space at some finite photon number (in our case 10) the steady state density matrix can be calculated and the time dependent Master equation can be solved by diagonalizing it numerically~\cite{Briegel93}. We use the quantum optics toolbox, a Matlab library written by Sze Tan~\cite{Tan99}, for this purpose. In the following, we adapt the formal notation
\[
\rho(t)=e^{\mathcal{L}t}\rho_0(0)
\]
referring to an explicit solution to the master equation with the initial state $\rho_0$.

\section{Quantum Regression Theorem}
With this solution to the master equation at hand, one can use the quantum regression theorem~\cite{Lax63} to evaluate time dependent correlation functions. For the second-order correlation function, it states
\[
\langle a^\dagger a^\dagger(\tau)a(\tau)a\rangle=\textrm{tr}\left\{a^\dagger ae^{\mathcal{L}\tau}\left[a\rho_{\textrm{ss}}a^\dagger\right]\right\}.
\]
Reading it from right to left, its interpretation is quite straightforward: $a\rho_{\textrm{ss}}a^\dagger$ is the density matrix of the system after a photon has been annihilated. The term $e^{\mathcal{L}\tau}\left[a\rho_{\textrm{ss}}a^\dagger\right]$ designates its time evolution. Except for the normalization, the correlation function thus measures the average photon number after the system has been perturbed by the detection of a photon.

The numerical calculation via the quantum optics toolbox simply follows the same steps: determine the steady-state density matrix  $\rho_{\textrm{ss}}$, evaluate $a\rho_{\textrm{ss}}a^\dagger$, propagate it in time by numerically solving the Master equation as discussed above, evaluate the expectation value of the average photon number at time $\tau$.

Similar arguments hold for the third-order correlation functions. Here, the quantum regression theorem reads
\[
\langle a^\dagger a^\dagger(\tau_1)a^\dagger(\tau_1+\tau_2)a(\tau_1+\tau_2)a(\tau_1)a\rangle=\textrm{tr}\left\{a^\dagger ae^{\mathcal{L}\tau_2}\left[ae^{\mathcal{L}\tau_1}\left[a\rho_{\textrm{ss}}a^\dagger\right]a^\dagger\right]\right\}.
\]
After the detection of the first photon the system is described by the density matrix $a\rho_{\textrm{ss}}a^\dagger$. It is propagated in time ($e^{\mathcal{L}\tau_1}\left[a\rho_{\textrm{ss}}a^\dagger\right]$) until a second photon is detected at $\tau_1$. The resulting density matrix is propagated again in time ($e^{\mathcal{L}\tau_2}\left[ae^{\mathcal{L}\tau_1}\left[a\rho_{\textrm{ss}}a^\dagger\right]a^\dagger\right]$). The correlation function is then proportional to the probability to detect a third photon at time $\tau_2$ after the detection of the previous photon pair.

\section{Position Averaging}
Our previous discussion enables us to calculate the correlation functions for a fixed coupling constant $g$ and fixed atom-cavity detuning $\Delta_a$. In practice, both parameters depend on the position of the atom and thus vary during the passage of the atom through the cavity mode.
We write $\langle a^\dagger a^\dagger (\tau) a (\tau) a \rangle_{\overrightarrow{x}}$ to denote the explicit position dependence of the correlation function. On timescales that are much shorter than the transit time of an atom through the cavity mode, the correlation function that we measure is then given by
\[
g^{(2)}(\tau)=\frac{\int{\langle a^\dagger a^\dagger(\tau)a(\tau)a\rangle_{\overrightarrow{x}}p(\overrightarrow{x})d\overrightarrow{x}}}{\left(\int{\langle a^\dagger a\rangle_{\overrightarrow{x}}p(\overrightarrow{x})d\overrightarrow{x}}\right)^2}
\]
where $p(\overrightarrow{x})$ is the probability to find an atom at the position $\overrightarrow{x}$. An analogous expression is valid for the third-order correlation function.

The integration has to be carried out over some finite volume around an anti-node of the cavity mode. The size of the volume depends on the transmission threshold that is used to detect the passing atoms. If the passage of the atoms through the cavity mode was unaffected by the near resonant probe light and the red-detuned stabilization light, one would expect a homogeneous distribution independent of position, i.e. $p(\overrightarrow{x})=p_0$. However, both laser fields cause an attractive potential \cite{Pinkse00} guiding the atom towards regions of strong coupling. Therefore, we increased iteratively the probability to find the atom at the cavity center by fine tuning the distribution function such that the correlation function matches the experimental data. This adjustment was made for one parameter set. The same distribution function was then used to reproduce all other experimental results shown in the paper using only minor variations of the integration volume and a vertical scaling factor as fit parameters.